\documentclass[twocolumn]{revtex4}

\usepackage{graphicx}
\usepackage{psfrag}
\usepackage{epsfig}
\usepackage{subfigure}
\usepackage{amssymb,amsmath}

\begin{document}

\title{Eliminating spurious velocities in the free energy lattice Boltzmann method}

\author{C. M. Pooley and  K. Furtado}

\affiliation{Rudolf Peierls Centre for Theoretical Physics, 1 Keble Road, Oxford, OX1 3NP, United Kingdom.}

\date{\today}

%--------------------------------------------------------------------------

\begin{abstract}
Spurious velocities are unphysical currents that appear close to curved interfaces in diffuse interface methods. We analyse the causes of these spurious velocities in the free energy lattice Boltzmann algorithm. By making a suitable choice of the equilibrium distribution, and by finding the best way to numerically calculate derivatives, we show that these velocities may be decreased by an order of magnitude compared to previous models. Furthermore, we propose a momentum conserving forcing method that reduces spurious velocities by another factor of $\sim 5$. In three dimensions we find that 19 velocity vectors is the minimum number necessary. 
\end{abstract}

\maketitle

\section{Introduction}
A commonly used approach for the simulation
of multi-phase fluid dynamics is the
free energy lattice Boltzmann method
introduced by Swift {\it et al.}~\cite{swift}.
This constitutes a so-called mesoscale
method because
it numerically solves the continuum equations of
fluid dynamics by exploiting the underlying
microscopic structure of these equations, without
resorting to a description of the fluid in terms
of molecular dynamics. One obstacle
to simulating some systems is that discretisation errors
lead to unphysical flows near interfaces.
These so-called spurious velocities
are present in multi-phase lattice
Boltzmann methods and in the other
diffuse interface methods.

An illustration of these spurious velocities is given in Fig. \ref{fig1}(a), which shows the flow profile around a liquid drop coexisting with a surrounding gas phase. 
The simulation is left until the long time steady state behaviour is reached. From a physical point of view all velocities should go to zero. What is observed, however, is that spurious flows persist indefinitely. 

A number of papers have dealt with this problem. Wagner \cite{wagner} analysed the case of binary fluids, and identified that one way to eradicate spurious velocities was to remove non-ideal terms from the pressure tensor and introduce these as a body force of the form ${\bf g} = -\phi \nabla \mu_\phi$. However, because this is no longer written in terms of the divergence of a pressure tensor (note that in general ${-\phi \nabla \mu_\phi}$ can always be rewritten as $-\partial_\beta P_{\alpha \beta}$) then momentum is no longer conserved.
Furthermore, Wagner pointed out that this method is numerically unstable unless some additional viscosity is artificially added to the system.

Lee and Fisher \cite{lee} use another forcing method for a different implementation of the lattice Boltzmann algorithm. Again, they eliminate spurious velocities at the expense of sacrificing momentum conservation. An additional difficulty with using forcing methods (including the one we present later) is that in order to update each lattice site the algorithm requires information from $2$ lattice sites away, rather than just $1$ for the standard method. This makes boundary conditions more complicated and slows down parallel computations, since more information needs to be passed between processors.  

Seta and Okui \cite{seta} used a lattice Boltzmann scheme proposed by Inamuro {\it et al.} \cite{inamuro}, and considered calculating the derivatives in the pressure tensor using a more accurate fourth-order scheme (as opposed to the usual second order accurate method). As will be shown later, however, they do not choose an optimum equilibrium distribution and hence their improvement in the spurious velocities is limited.
  
In this paper, we analyse the free energy lattice Boltzmann scheme for a liquid-gas system proposed by Swift {\it et al.} \cite{swift} and show that by making a careful choice of the equilibrium distribution (and also finding the best way to calculate derivatives) the magnitude of spurious velocities can be significantly reduced.
Furthermore, we present a second numerical scheme which moves gradient terms in the equilibrium distribution into a body force. This leads to a further reduction in spurious velocities whilst preserving momentum conservation.

The results of this analysis are equally applicable to other multiphase systems ({\it e.g.} binary fluids) when the free energy lattice Boltzmann method is used to solve their equations of motion.

\section{Model}

The pressure tensor for a liquid-gas system using a Landau free energy is given by
\begin{eqnarray}
P_{\alpha \beta} =  \left(p_0 - \kappa \rho \nabla^2 \rho - \frac \kappa 2 |\nabla \rho|^2 \right) \delta_{\alpha \beta} + \kappa \partial_{\alpha} \rho  \, \partial_{\beta} \rho,
\label{eq:pt}
\end{eqnarray}
where $\rho$ is the fluid density and $\kappa$ is a parameter related to the surface tension. We choose the bulk pressure $p_0$ to be that of a van der Waals fluid,
\begin{eqnarray}
p_0 = \frac{\rho T}{1-b \rho} - a\rho^2.
\label{p0}
\end{eqnarray}
This leads to liquid-gas phase separation below a critical temperature.

The analysis which follows is performed for a D2Q9 lattice Boltzmann scheme (in section \ref{3d} the results are summarised for the D3Q19 model) which uses a square lattice of side $\Delta x$, time-step $\Delta t$, and has $9$ velocity vectors, ${\bf e}_i$, where 
${\bf e}_0 = (0,0)$, ${\bf e}_{1,2} = (\pm c,0)$, ${\bf e}_{3,4} = (0,\pm c)$, ${\bf e}_{5,6} = (\pm c,\pm c)$, and ${\bf e}_{7,8} = (\mp c,\pm c)$. The parameter $c = \tfrac{\Delta x}{\Delta t}$ is a lattice velocity. 

A particle distribution function $f_i({\bf r},t)$ gives the mass density of particles travelling from lattice site ${\bf r}$, at time $t$, in a direction ${\bf e}_i$.
The physical variables are related to this distribution function by 
\begin{eqnarray}
\rho=\sum_i f_i,&\ \ \ \ \ \ &\rho u_{\alpha}=\sum_i f_i e_{i \alpha},
\label{rho}
\end{eqnarray}
where $\rho$ is the mass density and ${\bf u}$ is the velocity of the fluid.

The time evolution equation for the particle distribution function, using the standard BGK approximation, is given by
\begin{eqnarray}
 f_i({\bf r} + {\bf e}_i \Delta t , t+\Delta t) = f_i({\bf r}, t) - \tfrac{1}{\tau} \left[ f_i - f_i^{eq} \right]+ F_i,
\label{latbolt}
\end{eqnarray}
where $\tau$ is a relaxation parameter related to the viscosity, and $f^{eq}_i$ is an equilibrium distribution. It has been shown previously that this reduces to the Navier-Stokes equation provided the moments of $f_i^{eq}$ and $F_i$ are chosen suitably \cite{holdych} (see appendix \ref{app1}).
The final $F_i$ term is responsible for introducing a body force. This is not present in the standard formulation of the free energy lattice Boltzmann algorithm and so for now we set it to zero. In section \ref{force}, however, we discuss how this term can be usefully implemented to help reduce spurious velocities further.

The equilibrium distribution can be written as
\begin{eqnarray}
f^{eq}_i({\bf r}) &=& \tfrac{w_{i}}{c^{2}} \Big( e_{i\alpha} \rho u_{\alpha} 
+ \tfrac{3}{2c^{2}} \left[ e_{i\alpha} e_{i\beta} - \tfrac{c^{2}}{3}\delta_{\alpha\beta}\right] \times \nonumber\\
&& \quad \left( \rho u_\alpha u_\beta + \lambda \left[ u_\alpha \partial_\beta \rho + u_\beta \partial_\alpha \rho + \delta_{\alpha \beta} u_\gamma \partial_\gamma \rho \right] \right) \Big) \nonumber\\
&&\quad + \tfrac{1}{c^2} \Big( \,\, w_{i}^p p_0 - w_{i}^t \rho \nabla^2 \rho  + w_i^{xx} \kappa \partial_x \rho \partial_x \rho \nonumber\\
&&\quad + \,\, w_i^{yy} \kappa \partial_y \rho \partial_y \rho + w_i^{xy} \kappa \partial_x \rho \partial_y \rho \Big),
\label{equilibrium}
\end{eqnarray}
for $i=1,..,8$, where $w_{1\text{-}4} = \tfrac{1}{3}$, $w_{5\text{-}8} = \tfrac{1}{12}$, and summation over repeated indices is assumed.
The $i=0$ stationary value is chosen to conserve mass:
\begin{equation}
f_0^{eq}({\bf r}) = \rho - \sum_{i=1}^{8} f_i({\bf r}).
\label{eqn:feq0}
\end{equation}

The top two lines on the right hand side of Eq. \eqref{equilibrium} correspond to a standard expansion of the Maxwell Boltzmann distribution in discretised space \cite{luo}, and a correction term involving $\lambda$ (see Eq. (\ref{viscous})) which ensures Galilean invariance \cite{holdych}. 
%We do not write these terms in their most general form as they 
These terms are not important from the point of view of spurious velocities because they each contain the fluid velocity $u_\alpha$ to some power, which is expected to be zero in equilibrium. 

The last two lines in Eq. \eqref{equilibrium} give the pressure tensor contribution to the equilibrium distribution. This has been written in its most general form involving the free parameter weights $w_{i}^p$, $w_{i}^t$, $w_{i}^{xx}$, $w_{i}^{yy}$, and $w_{i}^{xy}$. Through the course of this paper optimum values for these parameters will be obtained.

The derivatives in the equilibrium distribution \eqref{equilibrium} are explicitly calculated within the algorithm using finite difference schemes. For instance, one simple choice for calculating the $x$ derivative of $\rho$ is given by
\begin{eqnarray}
\bar{\partial}_x  \rho = \tfrac{1}{2 \Delta x} \left[ \rho({\bf r} + {\bf e}_1) - \rho({\bf r} + {\bf e}_2) \right].
\label{drdx}
\end{eqnarray}
The bar above the partial derivative denotes that this is a discrete operator. By Taylor expanding the right hand side we find that 
\begin{eqnarray}
\bar{\partial}_x = \partial_x + \tfrac{1}{6} {\Delta x}^2 \partial^3_x + \dots
\end{eqnarray}
The discrete operator is correct up to second order but there are higher order terms which are responsible for generating the spurious flows.

A useful representation of finite difference operators is to denote them by stencils. For instance Eq. (\ref{drdx}) can be rewritten
\begin{eqnarray}
\bar{\partial_x} \rho  = \frac{1}{2\Delta x}
\left[
\begin{array}{ccc}
0 & 0 & 0 \\
-1 & 0 & 1 \\
0 & 0 & 0 \\
\end{array}
\right]_\rho.
\label{simple}
\end{eqnarray}
The central entry in the matrix represents the point at which the derivative is being made and the surrounding 8 entries correspond to the neighbouring lattice points surrounding this. This, however, is not the only choice for calculating the $x$ derivative. The most general stencil using only $9$ lattice nodes can be written
\begin{eqnarray}
\bar{\partial}_x  &=& 
\frac{1}{\Delta x}
\left[
\begin{array}{ccc}
-B & 0 & B \\
-A & 0 & A \\
-B & 0 & B \\
\end{array}
\right] \nonumber\\
&=& \partial_x + \tfrac{1}{6} {\Delta x} ^2 \partial^3_x + 2B {\Delta x}^2  \partial_y^2 \partial_x + \dots
\label{xderiv}
\end{eqnarray}
where $B$ is a free parameter which can be used to determine the third order term and $A$ is defined by $2A+4B = 1$.

Similarly, the Laplacian operator can be represented by
\begin{eqnarray}
\bar{\nabla}^2 \rho  &=& 
\frac{1}{{\Delta x}^2}
\left[
\begin{array}{ccc}
D & C & D \\
C & -4\left(C+D\right) & C \\
D & C & D \\
\end{array}
\right]_\rho \nonumber\\
&=& \nabla^2 + \tfrac{{\Delta x}^2}{12} \left(\partial_x^4 + \partial_y^4 \right) + D {\Delta x}^2 \partial_x^2 \partial_y^2 + \dots
\label{laplace}
\end{eqnarray}
where $C+2D = 1$.

In equilibrium the Navier-Stokes equation reduces to 
\begin{eqnarray}
0 = -\partial_\beta P_{\alpha \beta}. 
\label{main}
\end{eqnarray}
In terms of the lattice Boltzmann algorithm, the partial derivative operator acting on the pressure tensor in Eq. (\ref{main}) is implemented as a result of the choice of equilibrium distribution and the streaming and colliding operations. When $\tau = 1$ (in section \ref{taune} we discuss the more general case) the lattice Boltzmann equation (\ref{latbolt}) reduces to
\begin{eqnarray}
f_i({\bf r}  , t+\Delta t) = f_i^{eq}({\bf r} - {\bf e}_i \Delta t, t).
\end{eqnarray}

We consider the idealised case when at some time $t$ the system is at rest, {\it i.e.} ${\bf u}({\bf r},t) = 0$, and the density distribution is chosen such that the continuous operator equation (\ref{main}) is solved exactly. We ask the question what happens when the continuous operators are replaced by their discrete counterparts. In this case (\ref{main}) will no longer be exactly satisfied and instead there will be some spurious force ${\bf G}$ on the left hand side. 

Using Eq. (\ref{equilibrium}), this force can be expressed in terms of stencils of the various terms in the equilibrium distribution:
\begin{eqnarray}
%\bar{\partial}_\beta P_{x \beta}&=& 
G_x &=& \tfrac{1}{\Delta t} \left(\rho u_x({\bf r},t+\Delta t) - \rho u_x({\bf r},t) \right) \nonumber\\
&=& \tfrac{1}{\Delta t} \sum_i f_i({\bf r}, t+\Delta t) e_{ix} \nonumber\\
&=& \tfrac{1}{\Delta t} \sum_i f_i^{eq}({\bf r} -{\bf e}_i \Delta t , t) e_{ix} \nonumber\\
&=&
\! \frac{-1}{\Delta x} \left(
\left[
\begin{array}{ccc}
-w_{5\text{-}8}^p & 0 & w_{5\text{-}8}^p\\
-w_{1\text{-}4}^p & 0 & w_{1\text{-}4}^p\\
-w_{5\text{-}8}^p & 0 & w_{5\text{-}8}^p 
\end{array}
\right]_{p_0}
\!\!\!\!\!\!
-
\left[
\begin{array}{ccc}
-w_{5\text{-}8}^t & 0 & w_{5\text{-}8}^t\\
-w_{1\text{-}4}^t & 0 & w_{1\text{-}4}^t\\
-w_{5\text{-}8}^t & 0 & w_{5\text{-}8}^t 
\end{array}
\right]_{\kappa \rho \nabla^2 \rho}
\right.
\nonumber\\
&&
\hspace{0.6cm}
+
\left[
\begin{array}{ccc}
-w_{5\text{-}8}^{xx} & 0 & w_{5\text{-}8}^{xx}\\
-w_{1\text{-}2}^{xx} & 0 & w_{1\text{-}2}^{xx}\\
-w_{5\text{-}8}^{xx} & 0 & w_{5\text{-}8}^{xx}
\end{array}
\right]_{M_{xx}}
\!\!\!\!\!\!\!\!
+
\left[
\begin{array}{ccc}
-w_{5\text{-}8}^{yy} & 0 & w_{5\text{-}8}^{yy}\\
-w_{1\text{-}2}^{yy} & 0 & w_{1\text{-}2}^{yy}\\
-w_{5\text{-}8}^{yy} & 0 & w_{5\text{-}8}^{yy}
\end{array}
\right]_{M_{yy}}
\nonumber\\
&&
\hspace{0.6cm}
\left.
+
\left[
\begin{array}{ccc}
-w_{7\text{-}8}^{xy} & 0 & w_{5\text{-}6}^{xy} \\
-w_{1\text{-}4}^{xy} & 0 & w_{1\text{-}4}^{xy} \\
-w_{5\text{-}6}^{xy}  & 0 & w_{7\text{-}8}^{xy} 
\end{array}
\right]_{M_{xy}}
\right),
\label{expand2}
\end{eqnarray}
in, for example, the $x$ direction. Here, we define $M_{\alpha \beta} = \kappa \partial_\alpha \rho \partial_\beta \rho$.
In writing this we have made use of the symmetry properties of the system to immediately reduce the number of free parameters in the model. For instance, the bulk pressure $p_0$ does not have a preferred direction ({\it i.e.} it acts the same in the $x$ and $y$ directions) and hence we expect that $w_1^p=w_2^p=w_3^p=w_4^p$, which we denote by $w_{1\text{-}4}^p$, and $w_5^p=w_6^p=w_7^p=w_8^p=w_{5\text{-}8}^p$.
Other terms do have a preferred direction. For example, $M_{xx}$ is less restricted and has the constraints $w_1^{xx} = w_2^{xx}$, $w_3^{xx} = w_4^{xx}$ and $w_5^{xx} = w_6^{xx} = w_7^{xx} = w_8^{xx}$. Because the equilibrium should be invariant under simultaneous interchange of $x$ and $y$ and switching the velocities ${\bf e}_{1,2,7} \leftrightarrow {\bf e}_{3,4,8}$, then we expect that $w_{1\text{-}2}^{xx} = w_{3\text{-}4}^{yy}$, $w_{3\text{-}4}^{xx} = w_{1\text{-}2}^{yy}$, and $w_{5\text{-}8}^{xx} = w_{5\text{-}8}^{yy}$.

To first order, $G_x$ should agree with Eq. (\ref{main}), which in the $x$ direction is given by
\begin{eqnarray}
 G_x = - \partial_x \left( p_0 \!-\! \kappa \rho \nabla^2 \rho \right) \!-\! \tfrac{1}{2} \partial_x \left(M_{xx}\! -\! M_{yy} \right) \!-\! \partial_y M_{xy}.
\label{fi}
\end{eqnarray}
By comparing Eq. (\ref{fi}) with Eq. (\ref{expand2}) further restrictions are possible. 
For instance, by using Eq. (\ref{xderiv}), the  $p_0$ stencil becomes $-\partial_x p_0$ to second order provided that $2w_{1\text{-}4}^p + 4w_{5\text{-}8}^p = 1$.
Similarly, $2w_{1\text{-}4}^t + 4w_{5\text{-}8}^t = 1$, $2w_{1\text{-}2}^{xx} + 4w_{5\text{-}8}^{xx} = \tfrac{1}{2}$, $2w_{1\text{-}2}^{yy} + 4w_{5\text{-}8}^{yy} = -\tfrac{1}{2}$, $w_{1\text{-}4}^{xy} = 0$, $w_{5\text{-}6}^{xy} = \tfrac{1}{4}$, and $w_{7\text{-}8}^{xy} = -\tfrac{1}{4}$. These constraints are also necessary to obtain the correct moments of the equilibrium distribution in Eq. (\ref{pcons}).

Given all these conditions the spurious force can be rewritten as
\begin{eqnarray}
G_x &=&
\! \frac{-1}{\Delta x} \left(
\left[
\begin{array}{ccc}
-w_{5\text{-}8}^p & 0 & w_{5\text{-}8}^p\\
-\left(\tfrac{1}{2} - 2w_{5\text{-}8}^p\right) & 0 & \left(\tfrac{1}{2} - 2w_{5\text{-}8}^p\right)\\
-w_{5\text{-}8}^p & 0 & w_{5\text{-}8}^p 
\end{array}
\right]_{p_0}
\!\!\!\!\!\!
\right.
\nonumber\\
&&
\hspace{0.6cm}
-
\left[
\begin{array}{ccc}
-w_{5\text{-}8}^t & 0 & w_{5\text{-}8}^t\\
-\left(\tfrac{1}{2} - 2w_{5\text{-}8}^t\right) & 0 & \left(\tfrac{1}{2} - 2w_{5\text{-}8}^t\right)\\
-w_{5\text{-}8}^t & 0 & w_{5\text{-}8}^t 
\end{array}
\right]_{\kappa \rho \nabla^2 \rho}
\nonumber\\
&&
\hspace{0.6cm}
+
\left[
\begin{array}{ccc}
-w_{5\text{-}8}^{xx} & 0 & w_{5\text{-}8}^{xx}\\
-\left(\tfrac{1}{4} - 2w_{5\text{-}8}^{xx}\right) & 0 & \left(\tfrac{1}{4} - 2w_{5\text{-}8}^{xx}\right)\\
-w_{5\text{-}8}^{xx} & 0 & w_{5\text{-}8}^{xx}
\end{array}
\right]_{M_{xx}}
\!\!\!\!\!\!\!\!
\nonumber\\
&&
\hspace{0.6cm}
+
\left[
\begin{array}{ccc}
-w_{5\text{-}8}^{xx} & 0 & w_{5\text{-}8}^{xx}\\
-\left(-\tfrac{1}{4} - 2w_{5\text{-}8}^{xx}\right) & 0 & \left(-\tfrac{1}{4} - 2w_{5\text{-}8}^{xx}\right)\\
-w_{5\text{-}8}^{xx} & 0 & w_{5\text{-}8}^{xx}
\end{array}
\right]_{M_{yy}}
\nonumber\\
&&
\hspace{0.6cm}
\left.
+
\frac{1}{4}
\left[
\begin{array}{ccc}
1 & 0 & 1 \\
0 & 0 & 0 \\
-1  & 0 & -1 
\end{array}
\right]_{M_{xy}}
\right).
\label{expand}
\end{eqnarray}

There remains only three independent parameters in this expression, $w_{5\text{-}8}^p$, $w_{5\text{-}8}^t$, and $w_{5\text{-}8}^{xx}$. In the follow section we choose these unknowns in order to minimise the spurious velocity contribution.

\section{Determining a unique equilibrium distribution}
\label{unique}

In this section, we explicitly calculate the spurious force per unit volume ${\bf G}$ (see Eq. (\ref{expand}))
for the case of a liquid drop of radius $R$. If we take the origin to lie at the centre of the
drop, then the density $\rho$ is solely a function of distance from that origin $r = \sqrt{x^2+y^2}$. 
Taylor expanding the $p_0$ stencil (see Eq. (\ref{xderiv})) we find the contribution to the force from this term is given by  
\begin{eqnarray}
G_x^p = -\left(\partial_x + \tfrac{1}{6} {\Delta x}^2 \partial_x^3 + 2 w_{5\text{-}8}^p {\Delta x}^2  \partial_x \partial_y^2 \right) p_0.
\label{fxp}
\end{eqnarray}
Transforming from Cartesian into polar coordinates is achieved using the relations
\begin{eqnarray}
\partial_x \rightarrow \tfrac{x}{r} \partial_r, \quad \partial_y \rightarrow \tfrac{y}{r} \partial_r.
\end{eqnarray}
By sequentially substituting these operators and performing derivatives, Eq. (\ref{fxp}) can be rewritten 
\begin{eqnarray}
G_x^p &=&   -x D_r p_0 - x \left(\tfrac{1}{2} + 2 w_{5\text{-}8}^p\right) {\Delta x}^2 D_r^2 p_0 \nonumber\\
&& \hspace{0.5cm} - \left( \tfrac{1}{6} x^3 + 2 w_{5\text{-}8}^p x y^2 \right) {\Delta x}^2 D_r^3 p_0,
\end{eqnarray}
where we define $D_r = \tfrac{1}{r} \partial_r$. By symmetry, the $y$ component can be obtained by interchanging the $x$ and $y$ labels in this expression. 
The force can be decomposed into two terms; a term parallel and a term perpendicular to the interface.
The perpendicular contribution results in a small deviation in the Laplace pressure difference across the interface. The parallel term cannot be corrected for in this way
and thus it is responsible for inducing spurious flows. 

A parallel unit vector is given by ${\bf n}_\parallel = (-y,x)$ and, therefore, this tangential contribution can be calculated using
\begin{eqnarray}
{\bf G}^p.{\bf n}_\parallel = \left(x^3 y - x y^3 \right) \left( \tfrac{1}{6} - 2 w_{5\text{-}8}^p  \right) D_r^3 p_0.
\label{Fp}
\end{eqnarray}
This is zero provided that $w_{5\text{-}8}^p = \tfrac{1}{12}$.   
An analysis of other terms in Eq. (\ref{expand}) can be performed in a similar way. For instance, the force contribution from the Laplacian term is given by
\begin{eqnarray}
G^t_x &=& \left(\partial_x + \tfrac{1}{6} {\Delta x}^2 \partial_x^3 + 2 w_{5\text{-}8}^p {\Delta x}^2 \partial_x \partial_y^2 \right) 
\Big[ \kappa \rho \big( \nabla^2 \rho +  \nonumber\\
&& \tfrac{{\Delta x}^2}{12} \left(\partial_x^4 + \partial_y^4 \right) \rho + D {\Delta x}^2 \partial_x^2 \partial_y^2 \rho \big) \Big]. 
\end{eqnarray}

By repeating the process that was used to derive Eq. (\ref{Fp}), we find that this contribution vanishes provided that $w_{5\text{-}8}^p = \tfrac{1}{12}$ and $D = \tfrac{1}{6}$.

When transformed into polar coordinates, the tangential force from the $M_{\alpha \beta}$ terms in Eq. (\ref{expand}) is given by
\begin{eqnarray}
{\bf G}^M.{\bf n}_\parallel &=& \kappa \Big[\left(x^5 y - x y^5 \right) \left( -\tfrac{1}{12} - 2w_{5\text{-}8}^p  \right) {\Delta x}^2 D_r^3 (D_r \rho)^2   \nonumber\\
&&\hspace{-0.7cm}+  \left(x^3 y - x y^3 \right) \left( -\tfrac{1}{12} - w_{5\text{-}8}^p  \right) {\Delta x}^2 D_r^2 (D_r \rho)^2 \nonumber\\
&&\hspace{-0.7cm}+ \left(x^5 y - x y^5 \right) \left( \tfrac{1}{6} - 2B  \right) {\Delta x}^2 D_r \left[ (D_r \rho) (D_r^3 \rho) \right]\nonumber\\
&&\hspace{-0.7cm}+ \left(x^3 y - x y^3 \right) \left( \tfrac{1}{6} - 2B  \right) {\Delta x}^2 (D_r \rho) (D_r^3 \rho)   \Big]. 
\label{lines}
\end{eqnarray}

The last two lines on the right hand side become zero when $B = \tfrac{1}{12}$. Generally, it is not possible to make the first two lines simultaneously zero. However, it turns out that the first term dominates over the second, and so the best choice is $  w_{5\text{-}8}^p = -\tfrac{1}{24}$. The reason for this is that the width of the interface is much smaller than the radius of curvature. The density $\rho$ is approximately constant in the bulk regions but varies sharply in the interface. If we denote the width of the interface to be $W$ then the largest value for a derivative can be typically obtained using $\partial_r \sim \tfrac{1}{W}$. Since the operator $D_r$ appears one more time on the first line than the second, and it contains an extra factor of $x^2$ or $y^2$, then we expect the ratio in the magnitude of the first two lines to be approximately $\sim \tfrac{R^2}{WR} \sim \tfrac{R}{W}$. 
In fact, a detailed analysis explicitly calculating the two functions based on a hyperbolic tangent interface profile reveals that their maxima differ by a factor $3 \tfrac{R}{W}$. Thus provided $R \gg W$ the second line will be negligible compared to the first.

Now that we have obtained a unique choice for the equilibrium, it is interesting to note that, to the best of our knowledge, none of the previously proposed free energy lattice Boltzmann schemes make this optimum choice. For example, Inamuro {\it et al.} \cite{inamuro} choose $w_{5\text{-}8} = 0$ and Desplat {\it et al.} \cite{desplat} choose $w_{5\text{-}8} = -\tfrac{1}{72}$.

\section{Numerical results}

\begin{figure}
\begin{center}
\includegraphics[width = 3in]{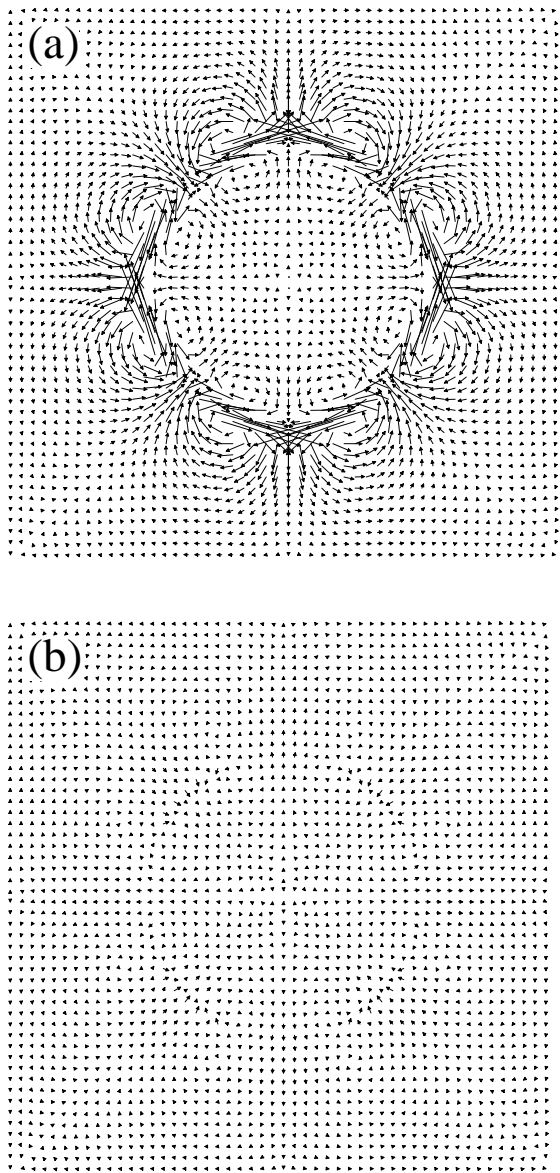}
\end{center}
\caption{The steady state velocity profile around a droplet using (a) a standard choice of equilibrium distribution and (b) an improved choice of equilibrium.}
\label{fig1}
\end{figure}

\begin{figure}
\begin{center}
\includegraphics[width = 3.2in]{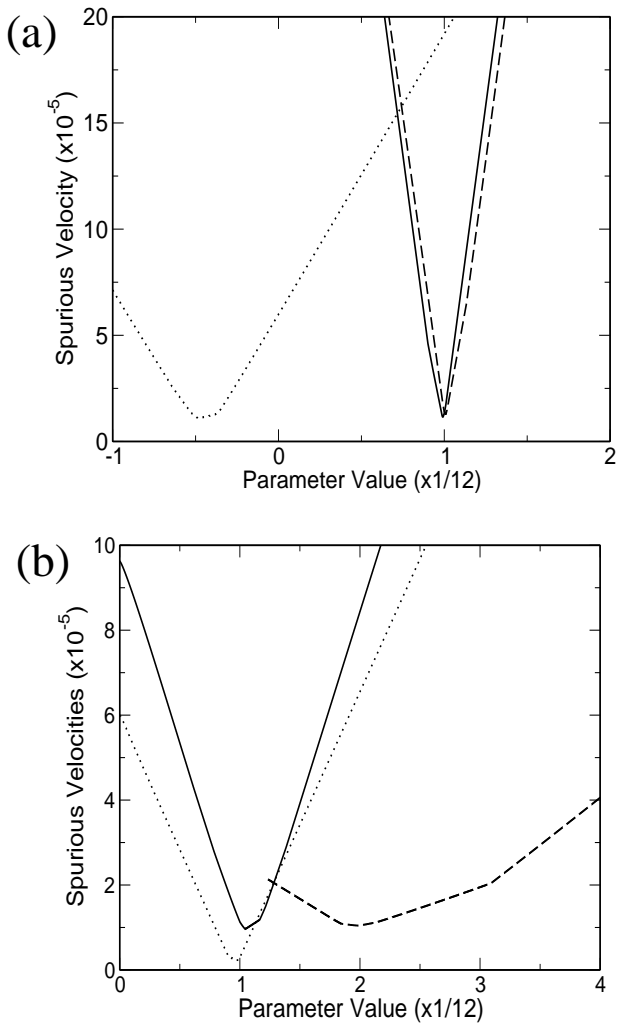}
\end{center}
\caption{The maximum spurious velocity as a function of (a) the equilibrium distribution parameters $w_{5\text{-}8}^p$ (solid line), $w_{5\text{-}8}^t$ (dashed line), $w_{5\text{-}8}^{xx}$ (dotted line), and (b) the stencil parameters $B$ (solid line), $D$ (dashed line), and the force stencil parameter $F$ (dotted line).}
\label{fig2}
\end{figure}

To test the predictions made in the previous section, we perform simulations on a grid of size $100 \times 100$. Parameters used were $a = \tfrac{9}{49}$, $b = \tfrac{2}{21}$, and $T = 0.56$, leading to liquid-gas phase separation with densities $\rho_l = 4.54$ and $\rho_g = 2.57$. The interfacial tension was set using $\kappa = 0.025$, giving an interface width of approximately $3$ lattice sites. 

A drop of radius $R=25$ was initialised at the centre of the system and simulations were run for $10^4$ time-steps to allow steady state conditions to be reached. 
Figure \ref{fig1}(a) shows the flow profile around the drop for a typical set of parameters. We clearly observe eight vortices in the gas phase surrounding the curved interface of the drop. Fig. \ref{fig1}(b) shows the dramatic reduction in the spurious flow when the best parameter choice is used. 

To verify that the we have, indeed, obtained an optimum choice of parameters, we show that the spurious velocity is minimised for each of the parameters separately. The effect of changing one parameter in isolation was found numerically by fixing all other degrees of freedom and scanning the chosen parameter's value over some range. This scanning procedure was performed sufficiently slowly to be quasi-static. Figure \ref{fig2} shows the results. The spurious velocity on the $y$-axis is defined to be the maximum velocity magnitude in the system. The solid curve in Fig. \ref{fig2}(a) shows how this velocity varies with $w_{5\text{-}8}^p$. It clearly reaches a minimum very close to that predicted theoretically ($w_{5\text{-}8}^p=\tfrac{1}{12}$). The spurious velocity never reaches exactly zero because our analysis only considered terms up to $O(\partial^4)$ in the Taylor series expansion for the stencils (Eqs. (\ref{xderiv}) and (\ref{laplace})). In reality, higher order terms also induce spurious velocities but these terms will be $\sim \tfrac{1}{W^2}$ smaller, and so have much less effect provided the interface width is reasonably large.

The other curves in Fig. \ref{fig2} show minima which correspond well with the values $w_{5\text{-}8}^t = \tfrac{1}{12}$, $w_{5\text{-}8}^{xx} = -\tfrac{1}{24}$, $B = \tfrac{1}{12}$, and $D = \tfrac{1}{6}$ predicted in section \ref{unique}. Note that when the simplest choice for calculating the derivatives is used (see Eq. (\ref{simple})), the spurious velocities are $\sim 10$ times larger than for the optimum choice (this corresponds to $B = 0$ in Fig. \ref{fig2}(b)).

\section{What happens when $\tau \neq 1$}
\label{taune}

To obtain Eq. (\ref{expand}) we assumed that $\tau = 1$ and so the lattice Boltzmann equation reduced to $f_i({\bf r}  , t+\Delta t) = f_i^{eq}({\bf r} - {\bf e}_i \Delta t, t)$. The more general case can be calculated under steady state conditions by sequentially substituting Eq. (\ref{latbolt}) back into the $f_i$ term on the right hand side. This gives 
\begin{eqnarray}
 f_i({\bf r}) &=& \tfrac{1}{\tau} \Big[ f_i^{eq}({\bf r} - {\bf e}_i \Delta t) + f_i^{eq}({\bf r} - 2{\bf e}_i \Delta t) \left(1-\tfrac{1}{\tau} \right) \nonumber\\
&&+  f_i^{eq}({\bf r}  - 3{\bf e}_i \Delta t) \left(1-\tfrac{1}{\tau} \right)^2 + \dots \Big].
\end{eqnarray}
Therefore, $f_i({\bf r})$ can be expressed in terms of the equilibrium distributions along lines of points radiating out following the velocity vector directions. The magnitude of these contributions decrease by a factor $z=(1-\tfrac{1}{\tau})$ for each step away. The stencils in Eq. (\ref{expand}) are no longer finite in size. For instance, the inner $5 \times 5$ region of the $p_0$ stencil now looks like
\begin{eqnarray}
\frac{1}{\tau} 
\left[
\begin{array}{ccccc}
-w_{5\text{-}8}^p z & 0 &  0 & 0 &  w_{5\text{-}8}^p z \\
0 & -w_{5\text{-}8}^p & 0 & w_{5\text{-}8}^p & 0\\
-w_{1\text{-}4}^p  & -w_{1\text{-}4}^p & 0 & w_{1\text{-}4}^p  &  w_{1\text{-}4}^p z\\
0 & -w_{5\text{-}8}^p & 0 & w_{5\text{-}8}^p  & 0\\
-w_{1\text{-}4}^p z & 0 &0 & 0 &  w_{5\text{-}8}^p z
\end{array}
\right]_{p_0}.
\end{eqnarray}
Converting this into continuous operators gives
\begin{eqnarray}
\bar{\partial}_x p_0 = \left(\partial_x \!+ \!S \left[ \tfrac{1}{6} {\Delta x}^2 \partial^3_x \!+\! 2w_{5\text{-}8}^p {\Delta x}^2  \partial_y^2 \partial_x \right] \!+\! \dots \right) p_0,
\end{eqnarray}
where the sum $S$ is 
\begin{eqnarray}
S &=& \frac{1}{\tau} \sum_{i=1}^{\infty} i^3 z^{i-1}\nonumber\\
&=& \tau - 6\tau^2 +6\tau^3.
\label{S}
\end{eqnarray}
Since $S$ is simply a numerical factor multiplying all the $O(\partial^3)$ terms then it will also pre-multiply the spurious force expressions in Eqs. (\ref{Fp}) and (\ref{lines}). Such a change does not alter the optimum choice of equilibrium when $\tau \neq 1$. 

\begin{figure}
\begin{center}
\includegraphics[width = 3in]{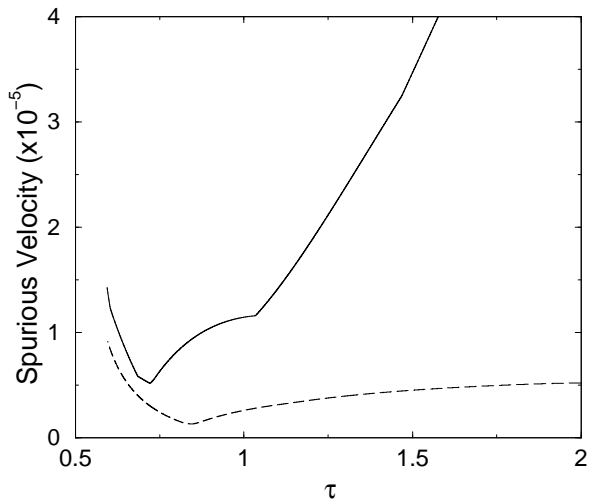}
\end{center}
\caption{The variation in spurious velocity as a function of $\tau$ for the standard LB (solid line) and for the new forcing method (dashed line).}
\label{fig3}
\end{figure}

The solid line in Fig. \ref{fig3} shows how the numerically calculated spurious velocities depend on $\tau$ using the optimum choice for all other parameters.
The function $S$ passes through zero when $\tau = \tfrac{1}{2} + \tfrac{1}{2\sqrt{3}} = 0.789$. This condition was calculated by Swift {\it et al.} \cite{swift} using a different method. It does not correspond exactly with the minimum of the curve because higher order spurious velocities become important in this region.

As $\tau$ is increased the spurious velocities rapidly increase in magnitude. These are principally generated by the small term on the second line of Eq. (\ref{lines}) being multiplied by the very large numerical factor $S$, which grows as $\tau^3$.

\section{Using a forcing method}
\label{force}

Rather than incorporate the problematic $M_{\alpha \beta}$ terms into the equilibrium distribution, it is also possible to put them into a body force.
The term $F_i$ in Eq. (\ref{latbolt}) is given by 
\begin{eqnarray}
F_i = \tfrac{w_{i}}{c^{2}} \left( e_{i\alpha} g_\alpha \!+\! \tfrac{3}{2c^{2}} \left[ e_{i\alpha} e_{i\beta} \!-\! \tfrac{c^{2}}{3} \delta_{\alpha\beta} \right] 
 \left(  u_\alpha g_\beta \!+ \!u_\beta g_\alpha \right)\right), \nonumber\\
\end{eqnarray}
where ${\bf g}$ is a body force that now appears on the right hand side of the Navier-Stokes equation (\ref{nsfinal2}). In this new forcing scheme, the $w_i^{xx}$, $w_i^{yy}$, and  $w_i^{xy}$ terms are removed from the equilibrium distribution (\ref{equilibrium}) and replaced by 
\begin{eqnarray}
g_x &=& - \tfrac{1}{2} \bar{\partial}_x \left(M_{xx} - M_{yy} \right) - \bar{\partial}_y M_{xy} \nonumber\\   
&=&
- \frac{1}{\Delta x} \left[
\begin{array}{ccc}
-F & 0 & F \\
-E & 0 & E \\
-F & 0 & F \\
\end{array}
\right]_{\tfrac{ M_{xx}-M_{yy}}{2}}
\nonumber\\ 
&&-
\frac{1}{\Delta x} 
\left[
\begin{array}{ccc}
F & E & F \\
0 & 0 & 0 \\
-F & -E & -F \\
\end{array}
\right]_{M_{xy}},
\end{eqnarray}
in the $x$ direction. $g_y$ may be obtained by interchanging the labels $x$ and $y$ and transposing the stencils.
Such a procedure leaves the continuum Navier-Stokes equation unchanged.

This method has the advantage of allowing extra degrees of freedom in choosing the stencils as compared to the standard lattice Boltzmann. In particular, we can choose to have a symmetry between the derivatives in the $x$ and $y$ directions ({\it i.e.} the $y$ stencil can be obtained by transposing the $x$). By comparison, Eq. (\ref{expand}) clearly cannot have this property. This improvement in the isotropy of the governing equation helps to reduce spurious velocities further.

The dotted line in Fig. \ref{fig2}(b) shows numerical results of how the spurious velocities change as a function of the stencil parameter $F$. The minimum of this curve lies at $F = \tfrac{1}{12}$, corresponding to a standard choice. By comparing the magnitude of the spurious velocity at this point with the minima from the other curves, we conclude that the forcing method leads to a further $\sim5$ fold reduction, giving a typical value of $\sim 2\times 10^{-6} c$.

Another advantage of using forcing is shown in Fig. \ref{fig3}. As $\tau$ is increased the spurious velocities normally become non-negligible due to the large numerical factor $S$ in Eq. (\ref{S}) multiplying the otherwise small contribution from the second line in Eq. (\ref{lines}). In the forcing method this term goes to zero allowing for accurate simulation of more viscous systems.

In general, the disadvantages of using forcing methods are that they make boundary conditions more complicated and, if being run on a parallel computer, require more information to be passed between computer micro-processors. This is because the standard two dimensional lattice Boltzmann method only requires information from the surrounding 8 points to update each lattice site, whereas the forcing method requires information from 24 points.    

\section{Extension to 3D lattice Boltzmann schemes}
\label{3d}

A number of different lattice Boltzmann schemes have been proposed for simulating 3D systems using $15$, $19$ or $27$ lattice velocities. In this paper we find that $19$ lattice vectors are necessary to ensure the reduction in spurious velocities. % We show the analysis for this case.
One way to define the velocity vectors in this model is the following: ${\bf e}_{1-6}$ lie along the nearest neighbour directions
\begin{eqnarray}
\left(
\begin{array}{c}
{e}_{x1\text{-}6} \\
{e}_{y1\text{-}6} \\
{e}_{z1\text{-}6}
\end{array}
\right)
=
\left[
\begin{array}{cccccc}
c & -c & 0 & 0 & 0 & 0\\
0 & 0 & c & -c & 0 & 0\\
0 & 0 & 0 & 0 & c & -c
\end{array} 
\right], \nonumber
\end{eqnarray}
and ${\bf e}_{7-18}$ are in the 12 square diagonal directions
\begin{eqnarray}
\left(
\begin{array}{c}
{e}_{x7\text{-}18} \\
{e}_{y7\text{-}18} \\
{e}_{z7\text{-}18}
\end{array}
\right)
\!=\!
\left[
\begin{array}{cccccccccccc}
c & \text{--}c& c  & \text{--}c & 0 & 0  & 0  &  0 & c & \text{--}c & c  & \text{--}c\\
c & +c & \text{--}c & \text{--}c & c & \text{--}c & c  & \text{--}c & 0 & 0  & 0  & 0\\
0 & 0  & 0  & 0  & c &  c & \text{--}c & \text{--}c & c & c  & \text{--}c & \text{--}c  
\end{array}
\right]\!. \nonumber
\end{eqnarray}

Analogous to the definitions for the gradient and Laplacian stencils given in Eqns. (\ref{xderiv}) and (\ref{laplace}), we define 
\begin{eqnarray}
\bar{\partial}_x \!\!&=& \!\!\frac{1}{\Delta x} \!\!
\left[ 
\!
\left(
\begin{array}{ccc}
 0 & 0 & 0 \\
-B & 0 & B \\
 0 & 0 & 0 \\
\end{array}
\right)
\!\!,\!\!
\left(
\begin{array}{ccc}
-B & 0 & B \\
-A & 0 & A \\
-B & 0 & B \\
\end{array}
\right)
\!\!,\!\!
\left(
\begin{array}{ccc}
 0 & 0 & 0 \\
-B & 0 & B \\
 0 & 0 & 0 \\
\end{array}
\right)
\! \right],
\nonumber\\
\bar{\nabla}^2\!\! &=&\!\! \frac{1}{\Delta x^2} \!\!
\left[ 
\!
\left(
\begin{array}{ccc}
 0 & D & 0 \\
 D & C & D \\
 0 & D & 0 \\
\end{array}
\right)
\!\!,\!\!
\left(
\begin{array}{ccc}
D & C & D \\
C & E & C \\
D & C & D \\
\end{array}
\right)
\!\!,\!\!
\left(
\begin{array}{ccc}
 0 & D & 0 \\
 D & C & D \\
 0 & D & 0 \\
\end{array}
\right)
\! \right], \nonumber\\
\label{laplacian3d}
\end{eqnarray}
where $E = -6C-12D$, $2A + 8B = 1$, $C+4D=1$ and the left, middle, and right matrices show slices of the stencil when $e_{zi} = c$,$0$, and $-c$, respectively.
 
In three dimensions, additional terms containing $w_i^{zz}$, $w_i^{zx}$, and $w_i^{yz}$ appear in the equilibrium distribution (\ref{equilibrium}).
By using the same procedure as in section \ref{unique}, this distribution can be uniquely defined. One additional complication in the three dimensional case is that there is no longer a single vector defining a tangent to the surface of a drop. Instead we use three vectors ${\bf n}^1_\parallel = (-y,x,0)$, ${\bf n}^2_\parallel = (0,-z,y)$, and ${\bf n}^3_\parallel = (-z,0,x)$ and require that the spurious force parallel to each one of these is zero. For instance, the previous expression in Eq. (\ref{lines}) becomes
\begin{eqnarray}
{\bf G}^M.{\bf n}^1_\parallel &=& \kappa \Big[  \left(x^5 y - x y^5 \right) \left( -\tfrac{1}{12} - 2w_{7\text{-}10}^{xx}\right) {\Delta x}^2 D_r^3 (D_r \rho)^2   \nonumber\\
&& \hspace{-2cm} + \left(yx^3z^2-xy^3z^2\right) \left(\tfrac{5}{12} + 2w_{7-10}^{xx} - 4w_{11-14}^{xx}\right)  {\Delta x}^2 D_r^3 (D_r \rho)^2  \nonumber\\
&& \hspace{-2cm} + \left(x^5 y - x y^5 \right) \left( \tfrac{1}{6} - 2B \right) {\Delta x}^2 D_r  (D_r \rho) (D_r^3 \rho) \nonumber\\
&&\hspace{-2cm}  + \left( yx^3z^2-xy^3z^2\right) \left( \tfrac{1}{6} - 2B  \right) {\Delta x}^2 D_r  (D_r \rho) (D_r^3 \rho)\Big].\nonumber\\
\end{eqnarray}
For the first line to be zero then $w_{7\text{-}10}^{xx} = -\tfrac{1}{24}$. The second line, therefore, is zero only when  $w_{11\text{-}14}^{xx} = \tfrac{1}{12}$. The last two lines vanish when $B = \tfrac{1}{12}$. 

A summary of all parameters obtained using this procedure is given below:
\begin{eqnarray}
w_{1\text{-}6} &=& w_{1\text{-}6}^p = w_{1\text{-}6}^t = \tfrac{1}{3}, \nonumber\\ 
w_{7\text{-}18} &=& w_{7\text{-}18}^p =  w_{7\text{-}18}^t  = \tfrac{1}{12},\nonumber\\
w_{1,2}^{xx} &=& w_{3,4}^{yy} = w_{5,6}^{zz} = \tfrac{1}{3},  \nonumber\\
w_{3\text{-}6}^{xx} &=& w_{1,2,5,6}^{yy} = w_{1\text{-}4}^{zz} = -\tfrac{1}{3},\nonumber\\
w_{7\text{-}10}^{xx} &=& w_{15\text{-}18}^{xx} = w_{7\text{-}14}^{yy} = w_{11\text{-}18}^{zz} = -\tfrac{1}{24}, \nonumber\\
w_{11\text{-}14}^{xx} &=& w_{15\text{-}18}^{yy} = w_{7\text{-}10}^{zz} = \tfrac{1}{12}, \nonumber\\
w_{1\text{-}6}^{xy} &=& w_{1\text{-}6}^{yz} = w_{1\text{-}6}^{zx} = 0,\nonumber\\  
w_{7,10}^{xy} &=& w_{11,14}^{yz} = w_{15,18}^{zx} = \tfrac{1}{4},\nonumber\\
w_{8,9}^{xy} &=& w_{12,13}^{yz} = w_{16,17}^{zx} = -\tfrac{1}{4},\nonumber\\
w_{11\text{-}18}^{xy} &=& w_{7\text{-}10}^{yz} = w_{15\text{-}18}^{yz} = w_{7\text{-}14}^{zx} = 0,\nonumber\\ 
&&\hspace{-1.5cm} A = \tfrac{1}{6}, \quad B  = \tfrac{1}{12}, \quad C = \tfrac{1}{3}, \quad D  = \tfrac{1}{6}.
\label{3dres}
\end{eqnarray}
Note that for a system one lattice unit wide this equilibrium reduces to the 2D result.

\section{Summary and conclusions}

In this paper we analysed the spurious velocities from two different methods: a standard lattice Boltzmann scheme and a new forcing method. 

Firstly, we calculated the spurious forces which originate when the continuous operators in the Navier-Stokes equation are replaced by stencils (in other word the contribution from the next order in the Taylor series expansion of the stencils).
Secondly, we identify that spurious velocities result from the component of these spurious forces acting parallel to the interface.
Finally, we find that by making a suitable choice of the equilibrium distribution and stencils we were able to set these parallel forces to zero (up to fourth order in the derivatives).

In 2D, the best choice of stencils for calculating the derivatives and the Laplacian are:
\begin{eqnarray}
\bar{\partial}_x  = 
\tfrac{1}{12 \Delta x}
\left[
\begin{array}{ccc}
-1 & 0 & 1 \\
-4 & 0 & 4 \\
-1 & 0 & 1 \\
\end{array}
\right]
, %\quad
\bar{\nabla}^2  = 
\tfrac{1}{6 {\Delta x}^2}
\left[
\begin{array}{ccc}
1 & 4 & 1 \\
4 & -20 & 4 \\
1 & 4 & 1 \\
\end{array}
\right].
\label{best}
\end{eqnarray}

Using the standard lattice Boltzmann model the equilibrium is given by Eq. (\ref{equilibrium}), where the optimum choice of parameters is 
$w_{1\text{-}4} = w_{1\text{-}4}^p = w_{1\text{-}4}^t = \frac{1}{3}$, 
$w_{5\text{-}8} = w_{5\text{-}8}^p = w_{5\text{-}8}^t = \frac{1}{12}$, 
$w_{5\text{-}8}^{xx} = w_{5\text{-}8}^{yy} = -\tfrac{1}{24}$,
$w_{1\text{-}2}^{xx} = w_{3\text{-}4}^{yy} = \tfrac{1}{3}$,
$w_{3\text{-}4}^{xx} = w_{1\text{-}2}^{yy} = -\tfrac{1}{6}$, $w_{1\text{-}4}^{xy} = 0$, and $w_{5\text{-}8}^{xy} = \tfrac{1}{4}$.
In 3D the corresponding results are summarised in Eqns. (\ref{laplacian3d}) and (\ref{3dres}).

One way to improve spurious velocities further is to remove the $M_{\alpha \beta} = \kappa \partial_\alpha \rho \partial_\beta \rho$ terms 
from the equilibrium distribution and implement them as a body force.
This force is then explicitly calculated by taking derivatives of $M_{\alpha \beta}$ using the stencil in Eq. (\ref{best}) (or Eq. (\ref{laplacian3d}) in the 3D case).
The additional symmetry in the resulting equations leads to a further reduction in spurious velocity size.

\acknowledgements

The authors would like to thank Prof. J.M. Yeomans for her help in writing this paper.

\appendix
\section{The moments}
\label{app1}

To conserve mass and momentum the first two constraints on the equilibrium distribution must be
\begin{eqnarray}
\sum_i f_i^{eq} = \rho,&\ \ \ \ \ \ 
&\sum_i f_i^{eq}e_{i \alpha}= \rho u_{\alpha}.
\label{ncons}
\end{eqnarray}
The higher order moments of $f_i^{eq}$ 
are chosen such that the resulting continuum equations
describe the dynamics of a non-ideal fluid. 
A suitable choice is
\begin{eqnarray}
\sum_i f_i^{eq} e_{i\alpha} e_{i\beta} &=& P_{\alpha \beta} + \rho u_\alpha u_\beta  + \lambda \left( u_\alpha \partial_\beta \rho \right.\nonumber\\
&& \hspace{0.1cm}                                    \left. + u_\beta \partial_\alpha \rho + \delta_{\alpha \beta} u_\gamma \partial_\gamma \rho \right) , \label{pcons}\\
\hspace{-0.5cm}
\sum_i f_i^{eq} e_{i\alpha} e_{i\beta} e_{i\gamma} &=& \frac{\rho c^2}{3}(\delta_{\alpha \beta} u_\gamma \! +\! \delta_{\alpha \gamma} u_\beta \!+\! \delta_{\beta \gamma} u_\alpha) \label{focons},
\end{eqnarray} 
where 
\begin{equation}
\nu= \frac{{\Delta x}^2(\tau -\frac{1}{2})}{3\Delta t}, \quad \quad \lambda = \nu \left(1-\frac{3 c_s^2}{c^2} \right)
\label{viscous} 
\end{equation}
will become the shear and bulk kinematic viscosities, respectively.
The speed of sound is given by $c_s^2 = \frac{d p_0}{d \rho}$, where $p_0$ is the fluid pressure (\ref{p0}).  
The term involving $\lambda$ on the right hand side of Eq. (\ref{pcons}) is necessary to ensure Galilean 
invariance \cite{holdych}. For an ideal gas with $c_s^2 = \frac{c^2}{3}$ it is zero, but in the more 
general case it must be included.

The moments of the forcing term $F_i$ are defined by 
\begin{eqnarray}
&&\sum_i F_i = 0, \quad \quad \quad \sum_i F_i e_{i\alpha} = \Delta t g_{\alpha},\nonumber\\
&&\hspace{0.5cm} \sum_i F_i e_{i\alpha} e_{i\beta} =  \Delta t \left( u_\alpha g_\beta + u_\beta g_\alpha \right),
\end{eqnarray}
where ${\bf g}$ is the body force per unit volume acting on the fluid.

By applying the Chapman-Enskog expansion to the lattice Boltzmann equation (\ref{latbolt}) \cite{luo}, we 
obtain the continuity equation for the total density
\begin{equation}
\partial_t \rho +\partial_\alpha( \rho v_\alpha) = 0 ,
\label{conteqn}
\end{equation}
and the Navier-Stokes equation for the fluid momentum
\begin{eqnarray}
  &&\partial_{t}(\rho v_\alpha) +  \partial_\beta ( \rho v_\alpha v_\beta) = -  \partial_\beta P_{\alpha \beta} + g_\beta\nonumber\\
  && \hspace{0.5cm}+ \partial_\beta  \left(  \nu \rho \left( \partial_\beta  v_\alpha + \partial_\alpha  v_\beta \right) + \lambda \rho \delta_{\alpha \beta} \partial_\gamma  v_\gamma  \right),
\label{nsfinal2}
\end{eqnarray}
where the fluid velocity is defined by
\begin{eqnarray}
{\bf v} = {\bf u} + \tfrac{\Delta t}{2 \rho} {\bf g}.
\end{eqnarray}
Note that this definition differs slightly from the lattice fluid velocity (\ref{rho}) in the case when the body force is non-zero. It is ${\bf v}$, and not ${\bf u}$, which is used to calculate the spurious velocities in section \ref{force}.

\end{document}